\documentclass[conference]{IEEEtran}
\usepackage{cite}

\ifCLASSINFOpdf
   \usepackage[pdftex]{graphicx}
   \graphicspath{{./pdf/}{./jpeg}/}
   \DeclareGraphicsExtensions{.pdf,.jpeg,.png}
\else
\fi
\usepackage[cmex10]{amsmath}
\usepackage{array}

\ifCLASSOPTIONcompsoc
  \usepackage[caption=false,font=normalsize,labelfont=sf,textfont=sf]{subfig}
\else
  \usepackage[caption=false,font=footnotesize]{subfig}
\fi
\usepackage{siunitx}
\usepackage{bm}
\usepackage{bbm}
\usepackage{amssymb}
\DeclareMathOperator*{\argmax}{arg\,max}

\usepackage{tikz}

\usetikzlibrary{decorations.pathreplacing}
\usetikzlibrary{shapes,arrows,fit,positioning}
\usetikzlibrary{plotmarks}
\usetikzlibrary{patterns}

\definecolor{blueTUD10}{rgb}{0.804,0.831,0.886}
\definecolor{blueTUD}{rgb}{0.043,0.165,0.318}
\definecolor{blueTUD50}{rgb}{0.388,0.424,0.557}

\usepackage{pgfplots}
\pgfplotsset{compat=newest}
\newlength\figureheight	
\newlength\figurewidth
\newcommand{\columnplot}{\setlength\figureheight{0.26\textwidth} \setlength\figurewidth{0.37\textwidth}}	

\tikzstyle{block} = [draw, fill=white, rectangle,minimum height=3em, minimum width=5.2em]	
\tikzstyle{block_rot} = [draw, fill=white, rectangle,minimum height=5.2em, minimum width=3em]
\tikzstyle{sum} = [draw, fill=white, circle, node distance=1em,path picture={\draw[black](path picture bounding box.south) -- (path picture bounding box.north) (path picture bounding box.west) -- (path picture bounding box.east);}]
\tikzstyle{coord} = [coordinate]

\tikzstyle{state}=[shape=circle,draw=black]	
\tikzstyle{lightedge}=[<-,dotted]
\tikzstyle{mainstate}=[state,thick]
\tikzstyle{mainedge}=[<-,thick]

\tikzstyle{symbol}=[shape=circle,draw=black,minimum width=1em,scale=0.6]	
\tikzstyle{sample1}=[shape=circle,draw,scale=0.3]	
\tikzstyle{sample2}=[shape=circle,draw,densely dashed,scale=0.3]	
\tikzstyle{interleave}=[shape=circle,draw,fill,minimum width=1em,scale=0.6]	

\tikzstyle{register} = [draw, fill=white, rectangle,minimum height=3em, minimum width=3em]	
\tikzstyle{mod2} = [draw, fill=white, circle, label={mod 2}, node distance=1em,path picture={\draw[black](path picture bounding box.south) -- (path picture bounding box.north) (path picture bounding box.west) -- (path picture bounding box.east);}]

\tikzset{>=latex}

\begin{document}
%
\title{Spectral Efficient Communications employing 1-Bit Quantization and Oversampling at the Receiver}

\author{\IEEEauthorblockN{Tim H{\"a}lsig, Lukas Landau, and Gerhard Fettweis}
\IEEEauthorblockA{Vodafone Chair Mobile Communications Systems\\
Technische Universit\"{a}t Dresden, 01062 Dresden, Germany\\
Email: \{tim.haelsig, lukas.landau, fettweis\}@tu-dresden.de}}



%


\maketitle

\begin{abstract}
To relax power consumption requirements in multi-gigabit/s communications systems low resolution quantization can be used. Information-theoretic results have shown that systems employing 1-bit quantization and oversampling are a viable option for this. This work investigates such a structure under the influence of additive Gaussian noise and two matched pulse shaping filters. It is described how a BCJR algorithm, based on a finite-state channel assumption, can be used to reconstruct symbols of higher order modulation schemes that allow the transmission of more than one bit per symbol. Furthermore, it is shown how symbol sources that are fitted to the 1-bit constraint can significantly improve the error rate performance of the system.
\end{abstract}

\let\thefootnote\relax\footnotetext{This work has been supported in part by the German Research Foundation in the framework of the Collaborative Research Center 912 "Highly Adaptive Energy-Efficient Computing" and by the European Social Fund in the framework of the Young Investigators Group "3D Chip-Stack Intraconnects".}

%
\IEEEpeerreviewmaketitle

\section{Introduction}
As data rates of communications systems scale towards multi-gigabit per second, the demands posed to the receiver, especially in terms of sampling rate, grow extensively. Finding suitable analog-to-digital converter (ADC) realizations for these systems becomes a challenge, as high sampling rates and resolutions are either unavailable or accompanied by high power consumptions \cite{Walden1999}. One approach to circumvent this problem is to use ADCs with reduced precision (e.g., 1-4 bits) \cite{Singh2009}, which is based on the fact that they can be designed easier and more efficient than their contemporaries with high precision. The most extreme case of reducing the resolution to one bit has been shown to provide promising information-theoretic results \cite{Singh2009_Limits}. It was furthermore shown that systems with such a strong quantization constraint can benefit strongly from oversampling \cite{Shamai1994},\cite{Krone2012_Comm}.

Prior work on this subject mostly considered information-theoretic investigations. In \cite{Koch2010} the authors proved that the capacity of a channel, which is lost due to 1-bit quantization, can be partially recovered by oversampling the received signal. It was additionally reported that it is possible to trade conversion resolution versus sampling rate to reduce the loss in capacity. Bit error rate (BER) optimal quantizers were investigated in \cite{Lu2010}. The presented approach allows a reduction of the required resolution and power consumption of the ADC, while maintaining performance of the regular uniform quantizer. It was shown in \cite{Krone2012_Comm} that a system with 1-bit quantization, which transmits quadrature amplitude modulated (QAM) signals, benefits strongly from oversampling in terms of achievable information rate. The authors also showed that the intuitive level of one bit per symbol can be exceeded by exploiting the noise or intersymbol interference (ISI) generated by the channel. This approach of exploiting the ISI of the channel was extended in \cite{Landau2013}. It was shown that this technique can make the full information rate of higher order modulation schemes (e.g., 16-QAM) accessible. 

This paper focuses on the implementation of a receiver concept that takes advantage of the ISI and whose performance can be measured in terms of error rates. It is therefore the main step for making the information-theoretic results that have been obtained in prior works accessible in a system design.

The rest of the paper is structured as follows:
Section \ref{sec:model} introduces the considered baseband system model. In Section \ref{sec:receiver} it is illustrated how the transmitted symbols can be reconstructed from the received symbols using a finite-state channel representation and a trellis-based algorithm. Section \ref{sec:sources} shows three different symbol source models and discusses their influence on the system. The numerical performance of different system configurations and their implications is examined in Section \ref{sec:results}. The final Section \ref{sec:conclusion} concludes the paper with a brief summary of the main results.

In this paper bold letters describe vectors, such as $\bm{y}_k$, and sequences of symbols are at times denoted as $x_{k-L}^k = x_{k-L},\ldots,x_k$ or $x_{}^k = x_{1},\ldots,x_k$.

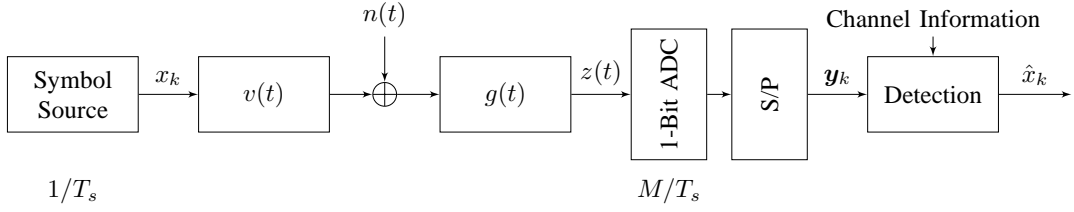
\begin{figure*}[!t]
\centering
\resizebox{0.8\textwidth}{!}{
\columnplot
\begin{tikzpicture}[auto, node distance=4em,>=latex']
    
		
		
		\node [block,align=center] (dac) {Symbol\\ Source};
    \node [block, right of=dac,node distance=7.6em] (vt) {$v(t)$};
		
		\node [sum,right of=vt,node distance=4.8em] (sum) {};
    \node [coord,label={$n(t)$},above of=sum,node distance=2.3em] (noise) {};
		
		\node [block, right of=sum,node distance=4.8em] (gt) {$g(t)$};

    \node [block_rot, right of=gt,node distance=6.5em] (adc) {\rotatebox{90}{1-Bit ADC}};
    \node [block_rot, right of=adc,node distance=4em] (sp) {\rotatebox{90}{S/P}};
    \node [block, right of=sp,node distance=6.5em] (equalizer) {Detection};	
		\node [coord,label={Channel Information},above of=equalizer,node distance=2.3em] (eqin) {};
	
    \node [coord,node distance=5.5em, right of=equalizer] (output) {};

		\node [coord,below of=dac,node distance=4.7em,label={$1/T_s$}] (signaling) {};
    \node [coord,below of=adc,node distance=4.7em,label={$M/T_s$}] (sampling) {};

    \draw [->] (dac) -- node {$x_k$} (vt);
    \draw [->] (vt) -- node {}(sum);
    \draw [->] (noise) -- node {}(sum);
    \draw [->] (sum) -- node {}(gt);
    \draw [->] (gt) -- node {$z(t)$}(adc);
    \draw [->] (adc) -- node {}(sp);
    \draw [->] (sp) -- node {$\bm{y}_k$}(equalizer);
		\draw [->] (eqin) -- node {}(equalizer);
		\draw [->] (equalizer) -- node {$\hat{x}_k$}(output);

\end{tikzpicture}}
\caption{Equivalent baseband model of a system that transmits data over an AWGN channel with matched filters, and uses oversampling and 1-bit conversion.}
\label{fig:system}
\end{figure*}

\section{System Model} \label{sec:model}
Assume an equivalent baseband system model of a communications system, as seen in Figure \ref{fig:system}. The input symbols $x_k$ are drawn from a finite set $\mathbb{X}$, which is determined by the chosen digital modulation scheme, at a symbol rate of $1/T_s$.
The two pulse shaping filters $v(t)$ and $g(t)$ can be chosen arbitrarily and are introduced to generate the necessary interference. We consider the noise $n(t)$ to have a Gaussian distributed amplitude and a constant spectral density. To create a fixed value for its variance, both the symbol energy and the pulse energies are normalized to one.

The received signal $z(t)$ can be written as
\begin{align}
z(t)=&\left(\sum_{k=-\infty}^{\infty}{x_k \delta(t-kT_s)} \ast v(t) + n(t) \right)\ast g(t) 
\text{.}
\end{align}
It is fed into the ADC where the signal is oversampled $M$-fold leading to $M$ samples per symbol duration $T_s$, which are stacked in the vector $\bm{y}_k=\left[y_{k,1},\ldots,y_{k,M}\right]$. The samples are furthermore processed by the 1-bit quantization operation $Q_1$ as
\begin{align}
y_{k,m} = Q_1 \left\{ z(t)\right\} = \begin{cases}  \phantom{-} 1 & \mbox{for } z(t)\geq 0 \\
-1 & \mbox{otherwise,} 
\end{cases}
\end{align}
where $m=1,\ldots,M$ and $t=kT_s + (m-1)T_s/M$. The quantized received vectors $\bm{y}_k$ are then fed into the block that tries to recover the input symbols $x_k$ based on the received symbol sequence $\bm{y}^n$ and channel state information.

We consider the channel to have finite-state (FS) property, where the current output symbol is influenced by the $L$ previous input symbols. This corresponds to the auxiliary channel assumption
\begin{align}
P\left(\bm{y}_k|\bm{y}^{k-1},x^k\right)\approx P\left(\bm{y}_k|x_{k-L}^k\right) 
\text{.}
\end{align}
It is assumed that $L$ is chosen such that the auxiliary channel sufficiently describes the principle property of the channel.

\section{Symbol Detection with BCJR}	\label{sec:receiver} 
We seek a method that exploits the FS property of the channel to reconstruct the transmitted symbols. This can be achieved with trellis-based algorithms, like maximum likelihood (ML) or maximum a posteriori (MAP), since the present FS channel can be represented by a trellis. We focus on MAP because it is a generalization of the ML approach which also considers the a posteriori probabilities of the states. The general maximum a posteriori detection rule can be written as
\begin{align}
\hat{x}_k = \argmax_{x_k \in \mathbb{X}}P(x_k|\bm{y}^n)\text{,}
\end{align}
where $x_k$ is a symbol at time $k$, $\mathbb{X}$ is the set of all possible symbols and $P(x_k|\bm{y}^n)$ is the probability for the symbol $x$ at time $k$ given the received sequence $\bm{y}^n = \bm{y}_1,\ldots,\bm{y}_n$.

\subsection{BCJR Algorithm}	%
The MAP approach can be efficiently implemented with the BCJR algorithm \cite{Bahl1974}. From the assumed channel model the states $s$ at time $k$ of the trellis are given by
\begin{align}
s_{k-1} = x_{k-L},\ldots,x_{k-1}	\label{eq:states}
\text{.}
\end{align}
The conventional decomposition of the APP $P(s_{k-1},s_{k},\bm{y}^{n})$ leads to the three parts that are required to carry out the algorithm. First the forward probability $\alpha$ containing all paths in the trellis leading to a state $s_k$ can be defined as 
\begin{align}
\alpha(s_{k})&=P(s_{k},\bm{y}_1,\ldots,\bm{y}_{k}) \notag \\
						&= \sum_{\forall s_{k-1}} \alpha_{} (s_{k-1}) \cdot \gamma_k (s_{k-1},s_k) 
\text{.}
\end{align}
With $\gamma_k$ being the probability for transitioning from state $s_{k-1}$ to $s_{k}$ given by
\begin{align}
\gamma_k(s_{k-1},s_k) &=P(s_k,\bm{y}_k|s_{k-1}) \notag \\
	&= P(s_{k}|s_{k-1})\cdot P(\bm{y}_k|s_{k},s_{k-1}) 
\text{.}
\end{align}
Furthermore, consider $\beta$ to be the backward probability as
\begin{align}
\beta(s_{k-1})&=P(\bm{y}_{k},\ldots,\bm{y}_{n}|s_{k-1}) \notag \\ 
						&= \sum_{\forall s_k} \beta_{} (s_k) \cdot \gamma_k (s_{k-1},s_k) 
\text{.}
\end{align}
It can then be concluded with 
\begin{align}
P(s_{k-1},s_k,\bm{y}^{n}) &=\alpha_{} (s_{k-1}) \cdot \gamma_{k} (s_{k-1},s_k) \cdot \beta_{} (s_k)
\text{,}
\end{align}
which leads to the APP using 
\begin{align}
P(s_{k-1},s_k|\bm{y}^{n}) = P(s_{k-1},s_k,\bm{y}^n)/P(\bm{y}^n)
\text{.}
\end{align}
Since $P(\bm{y}^n)$ is a constant for a given $\bm{y}^n$ the conditional probabilities can be obtained by normalizing the joint probabilities to add up to one. Note that it is necessary to find proper initial values for the recursions of $\alpha$ and $\beta$. Moreover, to efficiently implement the algorithm its matrix form can be used (e.g., \cite{Tuchler2011}).

Applying the state model from equation \eqref{eq:states} to the algorithm yields the following equations
\normalsize
\begin{align}
\gamma_k\left(x_{k-L}^{k}\right) &=P\left(x_{k}|x_{k-L}^{k-1}\right)\cdot P\left(\bm{y}_k|x_{k-L}^{k}\right) \notag \\
\alpha\left(x_{k-L+1}^{k}\right) &=\sum_{\forall x_{k-L}^{k-1}} \alpha_{} \left(x_{k-L}^{k-1}\right) \cdot \gamma_k \left(x_{k-L}^{k}\right)	\notag \\
\beta\left(x_{k-L}^{k-1}\right) &=\sum_{\forall x_{k-L+1}^{k}} \beta_{} \left(x_{k-L+1}^{k}\right) \cdot \gamma_k \left(x_{k-L}^{k}\right) \notag
\text{.}
\end{align}
\normalsize
The final a posteriori probability $P(x_k|\bm{y}^n)$, which is used for detection, can then be computed by summing over the branch APPs $P(s_{k-1},s_k|\bm{y}^n)$ that correspond to an input symbol of $x_k$ with
\small
\begin{align}
P(x_k|\bm{y}^n) &= \sum_{\forall s_{k-1},s_k\supseteq x_k} P(s_{k-1},s_k|\bm{y}^n)	\\
	&= \sum_{\forall x_{k-L}^k\supseteq x_k} P\left(x_{k-L}^k|\bm{y}^n\right)	\notag \\
	&= \sum_{\forall x_{k-L}^k\supseteq x_k} \alpha\left(x_{k-L}^{k-1}\right) \cdot \gamma_k\left(x_{k-L}^k\right) \cdot \beta\left(x_{k-L+1}^{k}\right) \notag
\text{.}
\end{align}
\normalsize

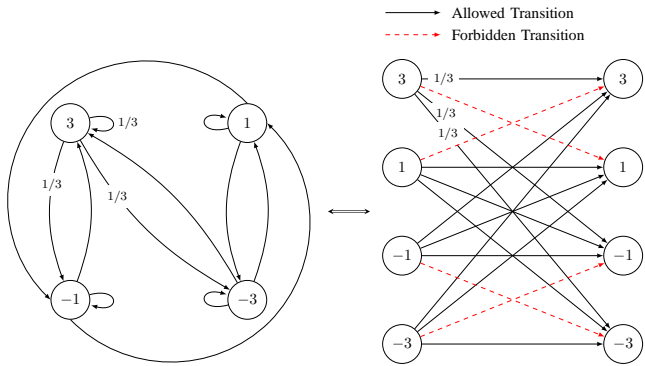
\begin{figure}[!b]
\centering
\resizebox{0.48\textwidth}{!}{\begin{tikzpicture}[]

\node[state,minimum width=2.5em] (s1) at (0,4) {$3$}
		edge [loop right] node{\footnotesize$1/3$} (t1);

\node[state,minimum width=2.5em] (s2) at (4,4) {$1$}
		edge [loop left] (t1);
\node[state,minimum width=2.5em] (s3) at (0,0) {$-1$}
		edge [loop right] (t1)
    edge [->,bend right=20] (s1)		
    edge [<-,bend left=20] node[fill=white,pos=0.7]{\footnotesize$1/3$} (s1);
\node[state,minimum width=2.5em] (s4) at (4,0) {$-3$}
		edge [loop left] (t1)
    edge [->,bend right=15] (s1)
    edge [<-,bend left=15] node[fill=white,pos=0.7]{\footnotesize$1/3$} (s1)
    edge [->,bend right=20] (s2)
    edge [<-,bend left=20] (s2);
 
\node[coordinate] (belows4) at (4.5,-0.5){}
    edge [-,bend left=45] (s3.south)
    edge [->,bend right=45] (s2.east);
    
\node[coordinate] (belows4) at (-0.5,4.5){}
    edge [->,bend right=45] (s3.west)
    edge [-,bend left=45] (s2.north);

\node (arrow1) at (5.7,2) {};
\node (arrow2) at (6.9,2) {};
\draw [<->,double,>=stealth] (arrow1) -- node {} (arrow2);
 
 \node[state,minimum width=2.5em] (state1_1) at (7.5,5) {$3$};
 \node[state,minimum width=2.5em] (state2_1) at (7.5,3) {$1$};
 \node[state,minimum width=2.5em] (state3_1) at (7.5,1) {$-1$};
 \node[state,minimum width=2.5em] (state4_1) at (7.5,-1) {$-3$};
 
 \node[state,minimum width=2.5em] (state1_2) at (12.5,5) {$3$};
 \node[state,minimum width=2.5em] (state2_2) at (12.5,3) {$1$};
 \node[state,minimum width=2.5em] (state3_2) at (12.5,1) {$-1$};
 \node[state,minimum width=2.5em] (state4_2) at (12.5,-1) {$-3$};

\draw [->] (state1_1) -- node[fill=white,pos=0.12] {\footnotesize$1/3$} (state1_2);
\draw [->] (state1_1) -- node[fill=white,pos=0.15] {\footnotesize$1/3$} (state3_2);
\draw [->] (state1_1) -- node[fill=white,pos=0.17] {\footnotesize$1/3$} (state4_2);

\draw [->] (state2_1) -- node {} (state2_2);
\draw [->] (state2_1) -- node {} (state3_2);
\draw [->] (state2_1) -- node {} (state4_2);

\draw [->] (state3_1) -- node {} (state1_2);
\draw [->] (state3_1) -- node {} (state2_2);
\draw [->] (state3_1) -- node {} (state3_2);

\draw [->] (state4_1) -- node {} (state1_2);
\draw [->] (state4_1) -- node {} (state2_2);
\draw [->] (state4_1) -- node {} (state4_2);

\draw [dashed,red,->] (state1_1) -- node {} (state2_2);
\draw [dashed,red,->] (state2_1) -- node {} (state1_2);
\draw [dashed,red,->] (state3_1) -- node {} (state4_2);
\draw [dashed,red,->] (state4_1) -- node {} (state3_2);

\node[] (legend1_1) at (7,6.5) {};
\node[] (legend1_2) at (8.5,6.5) {};

\node[] (legend2_1) at (7,6) {};
\node[] (legend2_2) at (8.5,6) {};

\node[anchor=west] (legend1) at (8.5,6.5) {Allowed Transition};
\node[anchor=west] (legend2) at (8.5,6) {Forbidden Transition};

\draw [->] (legend1_1) -- node[] {} (legend1_2);
\draw [dashed,red,->] (legend2_1) -- node[] {} (legend2_2);


\end{tikzpicture}}
\caption{A Markov source based on 4-ASK, state (left) and trellis (right) representation, avoiding symbol sequences that cannot be uniquely reconstructed. Corresponding transition probabilities are outlined on three example branches.}
\label{fig:markov}
\end{figure}

\subsection{Transition Probability}
For the algorithm to work it is necessary to identify what the term $P\left(x_{k}|x_{k-L}^{k-1}\right)\cdot P\left(\bm{y}_k|x_{k-L}^{k}\right)$ contains, as it resembles the ``Channel Information'' that is required prior to starting the algorithm. The first part of this term $P\left(x_{k}|x_{k-L}^{k-1}\right)$ accounts for the probability of the different symbol sequences. To give an example, for independent and uniformly distributed (i.u.d.) input variables this breaks down to $P(x_k)$, which is constant and can therefore be omitted when carrying out the algorithm. However, this has to be reassessed when the properties of the symbol source change.

The term $P\left(\bm{y}_k|x_{k-L}^{k}\right)$ is the probability that a certain output symbol $\bm{y}_k$ is observed given the sequence of input symbols $x_{k-L},\ldots,x_{k}$. There are two options to obtain this probability.

If there is knowledge about the appearance of the pulse shaping filter and the properties of the noise, it can be calculated by using a discrete representation of the system. This is done by integrating a multivariate Gaussian distribution over the appropriate quantization intervals. The distribution is characterized by the mean vector $\bm{\mu}_k$ given by
\begin{align}
\bm{\mu}_k & = \bm{H}\bm{U} x_{k-L}^{k} 
\text{,}
\end{align}
where $\bm{H}$ is the discrete filter matrix with Toeplitz structure of the combined pulse shape $h(t)=v(t)\ast g(t)$ and $\bm{U}$ is a upsampling matrix necessary for the discrete representation. The variance of the distribution is the noise covariance matrix $\bm{N}$ with 
\begin{align}
\bm{N} = \sigma_n^{2} \cdot \bm{G} \bm{G}^{\mathrm{H}} 
\text{,}
\end{align}
where $\bm{G}$ is the discrete filter matrix with Toeplitz structure of the receive pulse shaping filter $g(t)$.

The other possibility of obtaining $P\left(\bm{y}_k|x_{k-L}^{k}\right)$ is to estimate it. This requires that there is knowledge or at least an assumption about the channel memory $L$. A training sequence, which is known in the receiver, can then be sent through the system and in the receiver be used to form the relation between input sequence $x_{k-L},\ldots,x_k$ and output symbol $\bm{y}_k$.


\section{Symbol Sources} \label{sec:sources}
Due to the strong amplitude constraint given by the 1-bit quantizer, different symbol sources shall be considered that can be beneficial for the performance of the system. As the information about the symbol is, when using 1-bit quantization, given by the corresponding zero crossings of the signal, it is assumed that symbol sources generating more zero crossings could help to improve the performance.

\subsection{Independent and Uniformly Distributed Source}
A commonly used source is the one that draws symbols from a discrete alphabet with an independent and uniform distribution. It has the advantage of having the highest entropy of our considered sources. However, by using this source there is, using 1-bit ADC and BCJR detection, ambiguity when detecting certain sequences of symbols.

To give an example consider symbols drawn from a 4-ASK alphabet, which leads to an entropy of $2$ bits for the source. If there is now a sequence of ASK symbols where all of them have the same sign, the sequence does not generate any zero crossings. Therefore, the output sample sequence does not change no matter which of the two possible amplitude levels was actually sent, leading to a possible symbol error. 


\begin{figure}[!b]
\centering
\resizebox{0.48\textwidth}{!}{\begin{tikzpicture}[]

\node[state,minimum width=2.5em] (state1_1) at (0,3) {$3$};
 \node[state,minimum width=2.5em] (state2_1) at (0,1) {$1$};
 \node[state,minimum width=2.5em] (state3_1) at (0,-1) {$-1$};
 \node[state,minimum width=2.5em] (state4_1) at (0,-3) {$-3$};
 
 \node[state,minimum width=2.5em] (state1_2) at (4,3) {$3$};
 \node[state,minimum width=2.5em] (state2_2) at (4,1) {$1$};
 \node[state,minimum width=2.5em] (state3_2) at (4,-1) {$-1$};
 \node[state,minimum width=2.5em] (state4_2) at (4,-3) {$-3$};

\draw [->] (state2_1) -- node {} (state2_2);
\draw [->] (state1_1) -- node {} (state3_2);
\draw [->] (state2_1) -- node {} (state3_2);

\draw [->] (state2_1) -- node {} (state4_2);

\draw [->] (state3_1) -- node {} (state1_2);

\draw [->] (state3_1) -- node {} (state2_2);
\draw [->] (state4_1) -- node {} (state2_2);
\draw [->] (state3_1) -- node {} (state3_2);

\node[anchor=west] (legend1) at (-1,4.25) {Symbol Combinations for Super Symbol:};

\node[anchor=west] (legend2) at (7,4.25) {Signal Example with Rect Pulses, $M=3$:};


\node[coord] (threshold_start) at (6.75,0) {};
\node[coord] (threshold_end) at (14,0) {};
\draw[->,thick](threshold_start) -- (threshold_end);

\node[coord] (thresholdamp_start) at (6.75,-3.5) {};
\node[coord] (thresholdamp_end) at (6.75,3.5) {};
\draw[->,thick](thresholdamp_start) -- (thresholdamp_end);

\node[symbol] (symbol_1) at (8,1) {};
\node[symbol] (symbol_2) at (9.5,1) {};
\node[symbol] (symbol_3) at (11,1) {};
\node[symbol] (symbol_4) at (12.5,-3) {};

\draw[-](symbol_1) -- (symbol_2);
\draw[-](symbol_2) -- (symbol_3);
\draw[-](symbol_3) -- (symbol_4);

\node[coord] (sample1_1) at (7,0) {};
\node[sample2] (sample1_2) at (7,1) {};
\draw[dashed](sample1_1) -- (sample1_2);

\node[coord] (sample2_1) at (7.5,0) {};
\node[sample2] (sample2_2) at (7.5,1) {};
\draw[dashed](sample2_1) -- (sample2_2);

\node[coord] (sample3_1) at (8,0) {};
\node[sample1] (sample3_2) at (8,1) {};
\draw[](sample3_1) -- (sample3_2);

\node[coord] (sample4_1) at (8.5,0) {};
\node[sample2] (sample4_2) at (8.5,1) {};
\draw[dashed](sample4_1) -- (sample4_2);

\node[coord] (sample5_1) at (9,0) {};
\node[sample2] (sample5_2) at (9,1) {};
\draw[dashed](sample5_1) -- (sample5_2);

\node[coord] (sample6_1) at (9.5,0) {};
\node[sample1] (sample6_2) at (9.5,1) {};
\draw[](sample6_1) -- (sample6_2);

\node[coord] (sample7_1) at (10,0) {};
\node[sample2] (sample7_2) at (10,1) {};
\draw[dashed](sample7_1) -- (sample7_2);

\node[coord] (sample8_1) at (10.5,0) {};
\node[sample2] (sample8_2) at (10.5,1) {};
\draw[dashed](sample8_1) -- (sample8_2);

\node[coord] (sample9_1) at (11,0) {};
\node[sample1] (sample9_2) at (11,1) {};
\draw[](sample9_1) -- (sample9_2);

\node[coord] (sample10_1) at (11.5,0) {};
\node[sample2] (sample10_2) at (11.5,-1) {};
\draw[dashed](sample10_1) -- (sample10_2);

\node[coord] (sample11_1) at (12,0) {};
\node[sample2] (sample11_2) at (12,-1) {};
\draw[dashed](sample11_1) -- (sample11_2);

\node[coord] (sample12_1) at (12.5,0) {};
\node[sample1] (sample12_2) at (12.5,-1) {};
\draw[](sample12_1) -- (sample12_2);

\node[coord] (sample13_1) at (13,0) {};
\node[sample2] (sample13_2) at (13,-1) {};
\draw[dashed](sample13_1) -- (sample13_2);

\node[coord] (sample14_1) at (13.5,0) {};
\node[sample2] (sample14_2) at (13.5,-1) {};
\draw[dashed](sample14_1) -- (sample14_2);

\node[coord] (start) at (7,1) {};
\node[coord] (end) at (13.5,-1.33) {};

\draw[-](start) -- (symbol_1);
\draw[-](symbol_4) -- (end);

\node[coord,label={Time}] (lab2) at (13.5,0) {};
\node[coord,label={\rotatebox{90}{Amplitude}}] (lab_ampl) at (6.5,-1) {};

\draw[decorate,decoration={brace,amplitude=0.75em},xshift=0em,yshift=1.5em] (8,0.65) -- (9.5,0.65) node[black,midway,yshift=4mm,align=center]{\small Super Symbol};

\draw[decorate,decoration={brace,amplitude=0.75em,mirror},xshift=0em,yshift=1.5em] (7,-0.65) -- (10.5,-0.65) node[black,midway,yshift=-7.5mm,align=center]{\small Samples of\\ Super Symbol};

\node[draw=white, fill=white, rectangle] () at (6.5,-3.75) {};	

\end{tikzpicture}}
\caption{Possible super symbols constructed from 4-ASK (left) and a signal example with the corresponding sample assignment (right) that produces uniquely identifiable symbols.}
\label{fig:supersymbol}
\end{figure}
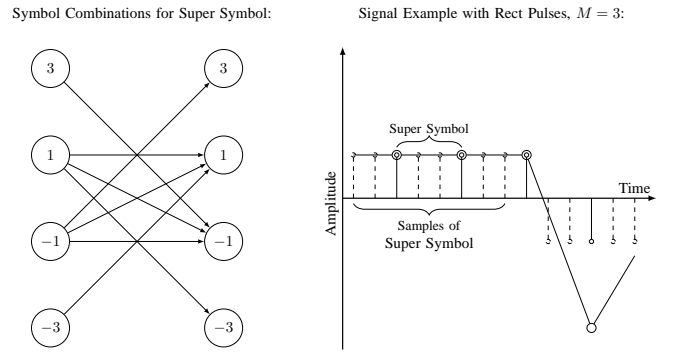

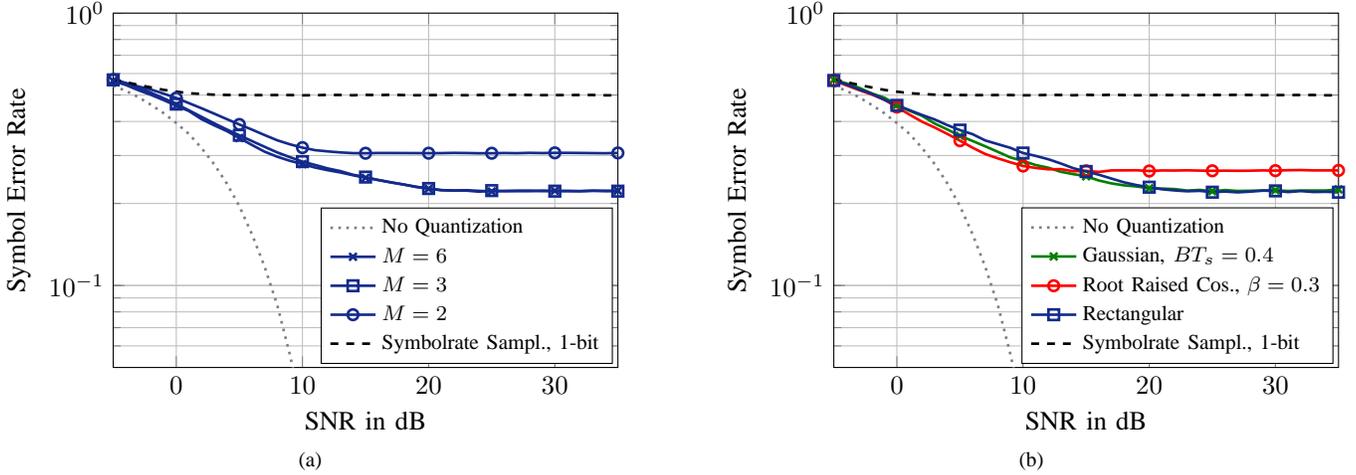
\begin{figure*}[!t]
\centering
\subfloat[]{\columnplot
%
%
%
\definecolor{mycolor1}{rgb}{0.501960813999176,0.501960813999176,0.501960813999176}%
\definecolor{mycolor2}{rgb}{0.0784313753247261,0.168627455830574,0.549019634723663}%
\begin{tikzpicture}

\begin{axis}[%
width=\figurewidth,
height=\figureheight,
scale only axis,
xmin=-5,
xmax=35,
xlabel={SNR in dB},
xmajorgrids,
ymode=log,
ymin=0.05,
ymax=1,
yminorticks=true,
ylabel={Symbol Error Rate},
ymajorgrids,
yminorgrids,
legend style={at={(0.99,0.01)},anchor=south east,draw=black,fill=white,legend cell align=left,font=\footnotesize}
]
\addplot [
color=mycolor1,
dotted,
line width=1.0pt
]
table[row sep=crcr]{
-5 0.541256\\
-4 0.518274\\
-3 0.491118\\
-2 0.460498\\
-1 0.42927\\
0 0.394796\\
1 0.36022\\
2 0.319784\\
3 0.278816\\
4 0.23695\\
5 0.19551\\
6 0.154264\\
7 0.11732\\
8 0.084782\\
9 0.055836\\
10 0.034364\\
11 0.018514\\
12 0.008738\\
13 0.00368\\
14 0.00115\\
15 0.000276\\
16 4e-05\\
17 4e-06\\
18 0\\
19 0\\
20 0\\
21 0\\
22 0\\
23 0\\
24 0\\
25 0\\
26 0\\
27 0\\
28 0\\
29 0\\
30 0\\
31 0\\
32 0\\
33 0\\
34 0\\
35 0\\
};
\addlegendentry{No Quantization};

\addplot [
color=mycolor2,
solid,
line width=1.0pt,
mark=x,
mark repeat={5},
mark options={solid}
]
table[row sep=crcr]{
-5 0.563730436\\
-4 0.544683455\\
-3 0.520975479\\
-2 0.500959499\\
-1 0.479212521\\
0 0.457\\
1 0.437\\
2 0.408\\
3 0.387427613\\
4 0.369\\
5 0.348\\
6 0.328\\
7 0.31\\
8 0.297\\
9 0.287\\
10 0.279\\
11 0.271\\
12 0.265\\
13 0.26\\
14 0.254\\
15 0.248476752\\
16 0.243861756\\
17 0.239289761\\
18 0.234727765\\
19 0.230511769\\
20 0.226592773\\
21 0.223824776\\
22 0.223209777\\
23 0.223646776\\
24 0.222555777\\
25 0.221464779\\
26 0.222429778\\
27 0.222936777\\
28 0.223\\
29 0.222055778\\
30 0.222605777\\
31 0.222566777\\
32 0.222003778\\
33 0.222526777\\
34 0.222825777\\
35 0.222832777\\
};
\addlegendentry{$M=6$};

\addplot [
color=mycolor2,
solid,
line width=1.0pt,
mark=square,
mark repeat={5},
mark options={solid}
]
table[row sep=crcr]{
-5 0.568679431\\
-4 0.551931448\\
-3 0.531478469\\
-2 0.51009649\\
-1 0.486716513\\
0 0.463329537\\
1 0.43978656\\
2 0.416\\
3 0.394\\
4 0.374\\
5 0.356\\
6 0.339\\
7 0.323\\
8 0.308\\
9 0.296\\
10 0.285\\
11 0.277\\
12 0.269\\
13 0.263\\
14 0.255\\
15 0.25\\
16 0.244\\
17 0.239\\
18 0.235\\
19 0.23\\
20 0.227\\
21 0.224\\
22 0.223\\
23 0.223\\
24 0.221\\
25 0.223\\
26 0.222\\
27 0.222\\
28 0.222\\
29 0.223\\
30 0.222\\
31 0.222\\
32 0.222\\
33 0.223\\
34 0.222\\
35 0.222\\
};
\addlegendentry{$M=3$};

\addplot [
color=mycolor2,
solid,
line width=1.0pt,
mark=o,
mark repeat={5},
mark options={solid}
]
table[row sep=crcr]{
-5 0.57041843\\
-4 0.553979446\\
-3 0.537673462\\
-2 0.523845476\\
-1 0.506291494\\
0 0.487805512\\
1 0.467162533\\
2 0.447341553\\
3 0.427544572\\
4 0.408217592\\
5 0.39012461\\
6 0.373458627\\
7 0.356681643\\
8 0.341277659\\
9 0.32988567\\
10 0.320760679\\
11 0.313902686\\
12 0.310712689\\
13 0.307985692\\
14 0.306126694\\
15 0.305838694\\
16 0.306485694\\
17 0.306498694\\
18 0.306325694\\
19 0.306192694\\
20 0.305987694\\
21 0.306150694\\
22 0.306777693\\
23 0.306390694\\
24 0.305770694\\
25 0.305861694\\
26 0.306453694\\
27 0.306006694\\
28 0.306875693\\
29 0.306299694\\
30 0.306871693\\
31 0.306864693\\
32 0.306584693\\
33 0.306191694\\
34 0.305866694\\
35 0.306626693\\
};
\addlegendentry{$M=2$};

\addplot [
color=black,
dashed,
line width=1.0pt
]
table[row sep=crcr]{
-5 0.572114428\\
-4 0.558082442\\
-3 0.544733455\\
-2 0.533438467\\
-1 0.522564477\\
0 0.514663485\\
1 0.508031492\\
2 0.504718495\\
3 0.501730498\\
4 0.500623499\\
5 0.5003535\\
6 0.4999065\\
7 0.4999565\\
8 0.5002415\\
9 0.5001755\\
10 0.499019501\\
11 0.4995245\\
12 0.500776499\\
13 0.5004935\\
14 0.499271501\\
15 0.500971499\\
16 0.4998975\\
17 0.501303499\\
18 0.5000845\\
19 0.499301501\\
20 0.4998575\\
21 0.5003345\\
22 0.4999025\\
23 0.499423501\\
24 0.5003575\\
25 0.500514499\\
26 0.5003985\\
27 0.501022499\\
28 0.500567499\\
29 0.4997635\\
30 0.5000195\\
31 0.5001265\\
32 0.5003755\\
33 0.5004275\\
34 0.499144501\\
35 0.499198501\\
};
\addlegendentry{Symbolrate Sampl., 1-bit};

\end{axis}
\end{tikzpicture}%
\label{fig:oversamp}}
\hfill
\subfloat[]{\columnplot
%
%
%
\definecolor{mycolor1}{rgb}{0.501960813999176,0.501960813999176,0.501960813999176}%
\definecolor{mycolor2}{rgb}{0.0784313753247261,0.168627455830574,0.549019634723663}%
\definecolor{mycolor3}{rgb}{0,0.498039215803146,0}%
\begin{tikzpicture}

\begin{axis}[%
width=\figurewidth,
height=\figureheight,
scale only axis,
xmin=-5,
xmax=35,
xlabel={SNR in dB},
xmajorgrids,
ymode=log,
ymin=0.05,
ymax=1,
yminorticks=true,
ylabel={Symbol Error Rate},
ymajorgrids,
yminorgrids,
legend style={at={(0.99,0.01)},anchor=south east,draw=black,fill=white,legend cell align=left,font=\footnotesize}
]
\addplot [
color=mycolor1,
dotted,
line width=1.0pt
]
table[row sep=crcr]{
-5 0.541256\\
-4 0.518274\\
-3 0.491118\\
-2 0.460498\\
-1 0.42927\\
0 0.394796\\
1 0.36022\\
2 0.319784\\
3 0.278816\\
4 0.23695\\
5 0.19551\\
6 0.154264\\
7 0.11732\\
8 0.084782\\
9 0.055836\\
10 0.034364\\
11 0.018514\\
12 0.008738\\
13 0.00368\\
14 0.00115\\
15 0.000276\\
16 4e-05\\
17 4e-06\\
18 0\\
19 0\\
20 0\\
21 0\\
22 0\\
23 0\\
24 0\\
25 0\\
26 0\\
27 0\\
28 0\\
29 0\\
30 0\\
31 0\\
32 0\\
33 0\\
34 0\\
35 0\\
};
\addlegendentry{No Quantization};

\addplot [
color=mycolor3,
solid,
line width=1.0pt,
mark=x,
mark repeat={5},
mark options={solid}
]
table[row sep=crcr]{
-5 0.5693\\
-4 0.55208\\
-3 0.53206\\
-2 0.50718\\
-1 0.48938\\
0 0.46028\\
1 0.43786\\
2 0.41958\\
3 0.39314\\
4 0.37396\\
5 0.3542\\
6 0.33922\\
7 0.3243\\
8 0.30958\\
9 0.29516\\
10 0.2866\\
11 0.27678\\
12 0.27062\\
13 0.2614\\
14 0.25622\\
15 0.25168\\
16 0.24294\\
17 0.23704\\
18 0.23314\\
19 0.23094\\
20 0.22748\\
21 0.22686\\
22 0.22592\\
23 0.2231\\
24 0.22432\\
25 0.22168\\
26 0.22406\\
27 0.22238\\
28 0.22268\\
29 0.22338\\
30 0.22358\\
31 0.22228\\
32 0.2232\\
33 0.22314\\
34 0.2232\\
35 0.22538\\
};
\addlegendentry{Gaussian, $BT_s=0.4$};

\addplot [
color=red,
solid,
line width=1.0pt,
mark=o,
mark repeat={5},
mark options={solid}
]
table[row sep=crcr]{
-5 0.5664\\
-4 0.54306\\
-3 0.52182\\
-2 0.50246\\
-1 0.48096\\
0 0.45192\\
1 0.42388\\
2 0.39916\\
3 0.3801\\
4 0.36004\\
5 0.34004\\
6 0.32048\\
7 0.3039\\
8 0.29294\\
9 0.28332\\
10 0.27456\\
11 0.27008\\
12 0.26704\\
13 0.26728\\
14 0.26268\\
15 0.26356\\
16 0.26116\\
17 0.26448\\
18 0.26464\\
19 0.2648\\
20 0.26316\\
21 0.26442\\
22 0.26424\\
23 0.26472\\
24 0.26408\\
25 0.26366\\
26 0.26396\\
27 0.2637\\
28 0.26386\\
29 0.26454\\
30 0.264\\
31 0.26498\\
32 0.26536\\
33 0.26452\\
34 0.26456\\
35 0.26454\\
};
\addlegendentry{Root Raised Cos., $\beta=0.3$};

\addplot [
color=mycolor2,
solid,
line width=1.0pt,
mark=square,
mark repeat={5},
mark options={solid}
]
table[row sep=crcr]{
-5 0.56582\\
-4 0.54764\\
-3 0.52416\\
-2 0.50616\\
-1 0.4809\\
0 0.45848\\
1 0.44032\\
2 0.4225\\
3 0.40684\\
4 0.38806\\
5 0.37194\\
6 0.35938\\
7 0.34086\\
8 0.33024\\
9 0.32012\\
10 0.30638\\
11 0.29904\\
12 0.29078\\
13 0.28036\\
14 0.26938\\
15 0.2621\\
16 0.25508\\
17 0.24694\\
18 0.23918\\
19 0.2336\\
20 0.22952\\
21 0.22556\\
22 0.22336\\
23 0.22176\\
24 0.22254\\
25 0.22004\\
26 0.2213\\
27 0.21976\\
28 0.22106\\
29 0.22288\\
30 0.2225\\
31 0.22032\\
32 0.22212\\
33 0.22042\\
34 0.22076\\
35 0.21998\\
};
\addlegendentry{Rectangular};

\addplot [
color=black,
dashed,
line width=1.0pt
]
table[row sep=crcr]{
-5 0.572114428\\
-4 0.558082442\\
-3 0.544733455\\
-2 0.533438467\\
-1 0.522564477\\
0 0.514663485\\
1 0.508031492\\
2 0.504718495\\
3 0.501730498\\
4 0.500623499\\
5 0.5003535\\
6 0.4999065\\
7 0.4999565\\
8 0.5002415\\
9 0.5001755\\
10 0.499019501\\
11 0.4995245\\
12 0.500776499\\
13 0.5004935\\
14 0.499271501\\
15 0.500971499\\
16 0.4998975\\
17 0.501303499\\
18 0.5000845\\
19 0.499301501\\
20 0.4998575\\
21 0.5003345\\
22 0.4999025\\
23 0.499423501\\
24 0.5003575\\
25 0.500514499\\
26 0.5003985\\
27 0.501022499\\
28 0.500567499\\
29 0.4997635\\
30 0.5000195\\
31 0.5001265\\
32 0.5003755\\
33 0.5004275\\
34 0.499144501\\
35 0.499198501\\
};
\addlegendentry{Symbolrate Sampl., 1-bit};

\end{axis}
\end{tikzpicture}%
\label{fig:pulses}}
\caption{Symbol error rates for 4-ASK i.u.d. input: \protect\subref{fig:oversamp} Matched rectangular pulses with different oversampling factors $M$; \protect\subref{fig:pulses} Different pulse shapes (roll-off $\beta$, 3-dB bandwidth-symbol time $BT_s$) with oversampling of $M=3$.}
\label{fig:ser_4ask}
\end{figure*}

\subsection{Markov State Model}
As we have seen in the previous example symbol errors can occur if there is ambiguity when reconstructing the symbols. One solution is therefore to use a source that forbids the transitions which lead to not uniquely identifiable sequences. This can be achieved with a Markov state model, which generally reduces entropy of the source but should improve error rate performance. Another thing that needs to be considered when using this source is the probability $P\left(x_{k}|x_{k-L}^{k-1}\right)$ used in the algorithm. Due to Markov property $P\left(x_{k}|x_{k-L}^{k-1}\right)=P(x_k|x_{k-1})$, which is either constant or zero and needs to be adhered to correspondingly in the algorithm.

To give an example, a possible Markov model for 4-ASK and $M=3$ can be seen in Figure \ref{fig:markov} (state and trellis representation). This source assures that if there are sequences of symbols with the same sign, their values can be uniquely determined based on the right starting symbol. Indeed, there are four symmetric Markov sources yielding the desired property. The one shown here was used as it has the lowest energy in higher frequency components. It is characterized by the states $x_j$, which are defined by the previous input symbol, and the state (symbol) transition, which is possible with probability of $P(x_{i}|x_j)=1/3$, given that the transition is not forbidden. For comparison if we were to include the red dashed lines (trellis representation) and change the transition probability to $1/4$, it would yield a conventional 4-ASK i.u.d. source.

The entropy of the source can be calculated by
\begin{align}
H(X_{\text{Markov}}) &= - \sum_{j=1}^4 \sum_{i=1}^4 P(x_j) P(x_{i}|x_j)\log_2 P(x_{i}|x_j) \\
	&\approx 1.585 \text{ bit}\notag
\text{}
\end{align}
where $P(x_j)=1/4$ is the stationary distribution of the state model. Instead of being able to transmit $2$ bits, as was the case for 4-ASK i.u.d., this Markov source can hence only transmit $1.585$ bits or more accurately, using $\log_3$, $1$ trit.

\subsection{Super Symbols}
Another way of ensuring possible error free detection of the symbols, while transmitting more than $1$ bit per symbol, is to use a source that transmits certain combinations of symbols, which can be uniquely detected by the receiver structure and shall be called super symbols.

With 4-ASK and $M=3$ it is for example possible to construct one super symbol from two regular symbols in a way that there are overall $8$ super symbols, which can even be detected in a super-symbol-by-super-symbol manner. One possible realization can be seen in Figure \ref{fig:supersymbol}. With eight possibilities per two symbols the entropy of the source results to $1.5$ bits.

It is possible to carry out a super-symbol-by-super-symbol detection because the sample combination, with oversampling $M=3$ and rectangular pulses, is for each super symbol unique. Detection with a BCJR algorithm adjusted to super symbols is viable as well.

\section{Numerical Results}	\label{sec:results}
The performance of different system configurations is going to be measured in terms of symbol error rates. For this, a sequence of $n$ symbols $x^n$ was transmitted through the system and the difference to the estimated sequence $\hat{x}^n$ was computed. For the results using the BCJR algorithm, the probability $P\left(\bm{y}_k|x_{k-L}^{k}\right)$ was estimated using a random pilot sequence of length $10^5$. This length assures, from our experience, that the difference to the actual calculated probability is negligibly small. It is always matched filtering assumed, meaning that $v(t)=g(t)$. Furthermore, the channel memory is fixed to one symbol with $L=1$, which still models the channel very good but decreases computational complexity significantly.

Note that we show symbol error rates because going towards a bit level transmission introduces a whole set of additional design tasks (e.g. mapping for the sources), which shall not be part of this work.

\subsection{4-ASK i.u.d. Input and BCJR Detection}
The results in Figure \ref{fig:ser_4ask} show the SER behavior of a system with a 4-ASK i.u.d. symbol source. A couple of different conclusions can be drawn from these results. First, with a structure employing 1-bit quantization, oversampling and FS channel assumption, it is possible to achieve a SER of $0.23$ when using 4-ASK. While this is not sufficient for a communications system, it shows that we can reconstruct more than $50\%$ of the symbols. 

Secondly, when considering rectangular pulses with duration $T_s$, an oversampling increase from $M=3$ to $M=6$ brings only marginal gain in the lower SNR region. Therefore, an oversampling factor of $M=3$ seems to be the most feasible choice for this system structure. It can furthermore be said that the influence of the pulse shape on the performance in the shown cases is rather small. However, the influence can be very strong, depending on the parameters of the pulse (i.e., roll-off, bandwidth).
The symbol errors that occur are present due to the fact that there are certain symbol sequences which cannot be uniquely recovered, as discussed earlier. 

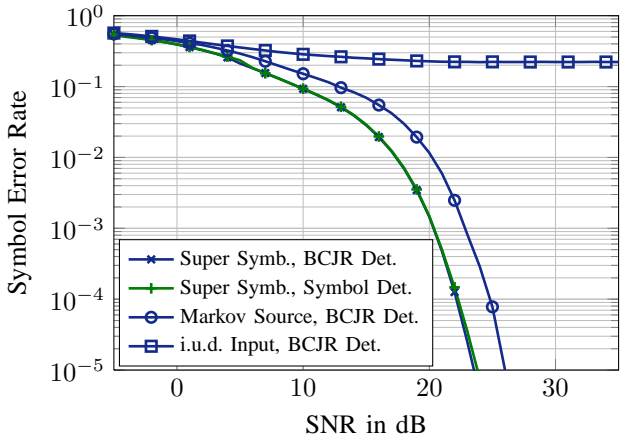
\begin{figure}[!t]
\centering
\columnplot
%
%
%
\definecolor{mycolor1}{rgb}{0.501960813999176,0.501960813999176,0.501960813999176}%
\definecolor{mycolor2}{rgb}{0.0784313753247261,0.168627455830574,0.549019634723663}%
\definecolor{mycolor3}{rgb}{0,0.498039215803146,0}%
\begin{tikzpicture}

\begin{axis}[%
width=\figurewidth,
height=\figureheight,
scale only axis,
xmin=-5,
xmax=35,
xlabel={SNR in dB},
xmajorgrids,
ymode=log,
ymin=1e-05,
ymax=1,
yminorticks=true,
ylabel={Symbol Error Rate},
ymajorgrids,
yminorgrids,
legend style={at={(0.01,0.01)},anchor=south west,draw=black,fill=white,legend cell align=left,font=\footnotesize} 
]


\addplot[
color=mycolor2,
solid,
line width=1.0pt,
mark=x,
mark repeat={3},
mark options={solid}
]
table[row sep=crcr]{
-5 0.540517919\\
-4 0.513310973\\
-3 0.484190032\\
-2 0.454689591\\
-1 0.424107652\\
0 0.392682715\\
1 0.359523781\\
2 0.325776848\\
3 0.293057414\\
4 0.257898984\\
5 0.211646577\\
6 0.181229638\\
7 0.15491069\\
8 0.131753736\\
9 0.110767278\\
10 0.092803314\\
11 0.077426845\\
12 0.063785872\\
13 0.051041898\\
14 0.039199422\\
15 0.028466943\\
16 0.019406961\\
17 0.012226476\\
18 0.006917986\\
19 0.003454493\\
20 0.001448997\\
21 0.000475499\\
22 0.000126\\
23 2.55e-05\\
24 4.5e-06\\
25 0\\
26 0\\
27 0\\
28 0\\
29 0\\
30 0\\
31 0\\
32 0\\
33 0\\
34 0\\
35 0\\
};
\addlegendentry{Super Symb., BCJR Det.};

\addplot [
color=mycolor3,
solid,
line width=1.0pt,
mark=+,
mark repeat={3},
mark options={solid}
]
table[row sep=crcr]{
-5 0.535323429\\
-4 0.502089496\\
-3 0.477996544\\
-2 0.451560597\\
-1 0.423642653\\
0 0.393630713\\
1 0.361982276\\
2 0.329402341\\
3 0.29481991\\
4 0.264317471\\
5 0.234482531\\
6 0.182458135\\
7 0.154054192\\
8 0.130982738\\
9 0.110567779\\
10 0.093299813\\
11 0.078097344\\
12 0.064583871\\
13 0.051820396\\
14 0.03983392\\
15 0.028987442\\
16 0.01975796\\
17 0.012491975\\
18 0.007105486\\
19 0.003536993\\
20 0.001487997\\
21 0.000500999\\
22 0.000147\\
23 3.6e-05\\
24 7.5e-06\\
25 0\\
26 0\\
27 0\\
28 0\\
29 0\\
30 0\\
31 0\\
32 0\\
33 0\\
34 0\\
35 0\\
};
\addlegendentry{Super Symb., Symbol Det.}; 

\addplot [
color=mycolor2,
solid,
line width=1.0pt,
mark=o,
mark repeat={3},
mark options={solid}
]
table[row sep=crcr]{
-5 0.552296895\\
-4 0.531050938\\
-3 0.50994898\\
-2 0.487105026\\
-1 0.464015072\\
0 0.436691127\\
1 0.410567179\\
2 0.385937228\\
3 0.353937292\\
4 0.322663355\\
5 0.291467417\\
6 0.259421481\\
7 0.227553545\\
8 0.198375603\\
9 0.172813654\\
10 0.151633697\\
11 0.132479735\\
12 0.11514377\\
13 0.096503807\\
14 0.083401833\\
15 0.069489861\\
16 0.05481389\\
17 0.042157916\\
18 0.02976194\\
19 0.019353961\\
20 0.011589977\\
21 0.005949988\\
22 0.002483995\\
23 0.000831998\\
24 0.000287999\\
25 7.79999e-05\\
26 9.99999e-06\\
27 0\\
28 0\\
29 0\\
30 0\\
31 0\\
32 0\\
33 0\\
34 0\\
35 0\\
};
\addlegendentry{Markov Source, BCJR Det.};

\addplot [
color=mycolor2,
solid,
line width=1.0pt,
mark=square,
mark repeat={3},
mark options={solid}
]
table[row sep=crcr]{
-5 0.568679431\\
-4 0.551931448\\
-3 0.531478469\\
-2 0.51009649\\
-1 0.486716513\\
0 0.463329537\\
1 0.43978656\\
2 0.416\\
3 0.394\\
4 0.374\\
5 0.356\\
6 0.339\\
7 0.323\\
8 0.308\\
9 0.296\\
10 0.285\\
11 0.277\\
12 0.269\\
13 0.263\\
14 0.255\\
15 0.25\\
16 0.244\\
17 0.239\\
18 0.235\\
19 0.23\\
20 0.227\\
21 0.224\\
22 0.223\\
23 0.223\\
24 0.221\\
25 0.223\\
26 0.222\\
27 0.222\\
28 0.222\\
29 0.223\\
30 0.222\\
31 0.222\\
32 0.222\\
33 0.223\\
34 0.222\\
35 0.222\\
};
\addlegendentry{i.u.d. Input, BCJR Det.};

\end{axis}
\end{tikzpicture}%
\caption{Symbol error rates for rectangular pulses (duration $T_s$), oversampling $M=3$ and different input sources.}
\label{fig:sources}
\end{figure}

\subsection{Markov Source and Super Symbols}
When applying the symbol sources from Section \ref{sec:sources} the error rates improve drastically and indeed tend towards zero, as can be seen in Figure \ref{fig:sources}. The results were computed with rectangular pulses of duration $T_s$ and oversampling of $M=3$.

For the Markov source results the model given in Figure \ref{fig:markov} was used and the corresponding BCJR detection algorithm was employed. The results look promising but two things have to be kept in mind. First when applying this source to the system the rate is reduced from $2$ bits per symbol, which is available with 4-ASK i.u.d., to about $1.58$ bits per symbol. Secondly mapping and demapping bits to this Markov source poses some problems regarding efficiency and error sensitivity.

For the super symbol results detection was done with a symbol-by-symbol approach as well as a MAP approach using the BCJR algorithm. For the symbol-by-symbol method the $8$ samples of the super symbol (as seen in Figure \ref{fig:supersymbol} on the right) were considered and the detected super symbol was chosen to be the one that is most likely to have produced this set of samples. It is also sufficient for unique reconstruction to just consider the $4$ samples in the center of the super symbol, but the SNR performance turned out to be slightly worse. The BCJR algorithm was used with the same assumptions as in the previous section but on a super symbol level. There are numerous possibilities when choosing the two symbols that constitute one super symbol, but the one shown in Figure \ref{fig:supersymbol} proved to have the best SNR performance.

The results show that the performance gain from using the more elaborate BCJR detection algorithm compared to the symbol-by-symbol approach is very small. This can be explained by the fact that the interference between two super symbols is, when using rectangular pulses, also very small. As for the Markov source, the entropy is reduced for this source as well, but mapping and demapping can be done very easily by assigning bit blocks to the super symbols.

Similar results can be computed with the two other pulse shapes as well (depending on the pulse parameters).

\section{Conclusion} \label{sec:conclusion}
In this paper a communications system with pulse shaping filters, oversampling and 1-bit quantization has been considered. It was explored how a finite-state channel assumption can be used to design a BCJR algorithm that allows the recovery of more than $50\%$ of 4-ASK symbols with this structure. However, the numerical results reveal that the overall error rate is too high for a practical implementation. Therefore, two alternative symbol sources having efficiencies of more than one bit per symbol were introduced that are more suitable for the given scenario. The results show that both are applicable, with SERs tending towards zero, but come at the cost of reduced rates compared to the standard i.u.d. source. Choosing fitting symbol sources and alphabets is consequently one of the easiest ways to design an efficient communications system with 1-bit quantization constraint.








%

\bibliographystyle{IEEEtran}
\bibliography{refs}

%
%

\end{document}